\documentclass[aps,pre,twocolumn,superscriptaddress]{revtex4}
\usepackage{graphicx}
\usepackage{amsmath}
\usepackage{multirow}

\begin{document} 

\title{Kinetics of double stranded DNA overstretching revealed by 0.5-2 pN
force steps} 

\author{Pasquale Bianco}
\affiliation{Laboratorio di Fisiologia, Dipartimento di Biologia Evoluzionistica, Universit\`a degli Studi di
Firenze, Via G. Sansone 1, I-50019 Sesto Fiorentino, Italy}
\author{Lorenzo Bongini}
\affiliation{Departament de Fisica Fonamental,
Universitat de Barcelona, Carrer Marti i Franqu\`es, 1 E-08028 Barcelona, Spain}, 
\author{Luca Melli}
\affiliation{Laboratorio di Fisiologia, Dipartimento di Biologia Evoluzionistica, Universit\`a degli Studi di
Firenze, Via G. Sansone 1, I-50019 Sesto Fiorentino, Italy}
\author{Mario Dolfi}
\affiliation{Laboratorio di Fisiologia, Dipartimento di Biologia Evoluzionistica, Universit\`a degli Studi di
Firenze, Via G. Sansone 1, I-50019 Sesto Fiorentino, Italy} 
\author{Vincenzo Lombardi}
\affiliation{Laboratorio di Fisiologia, Dipartimento di Biologia Evoluzionistica, Universit\`a degli Studi di
Firenze, Via G. Sansone 1, I-50019 Sesto Fiorentino, Italy}


\begin{abstract}
A detailed description of the conformational plasticity of double stranded DNA (ds) is a necessary
framework for understanding protein-DNA interactions. Until now, however structure and kinetics
of the transition from the basic conformation of ds-DNA (B state) to the 1.7 times longer and
partially unwound conformation (S state) have not been defined. The force-extension relation of the
ds-DNA of $\lambda$-phage is measured here with unprecedented resolution using a dual laser optical
tweezers that can impose millisecond force steps of 0.5-2 pN (25 $^{\circ}$C). This approach reveals the
kinetics of the transition between intermediate states of ds-DNA and uncovers the load-dependence
of the rate constant of the unitary reaction step. DNA overstretching transition results essentially a
two-state reaction composed of 5.85 nm steps, indicating cooperativity of ~25 base pairs. This
mechanism increases the free energy for the unitary reaction to ~94 $k_BT$, accounting for the stability
of the basic conformation of DNA, and explains the absence of hysteresis in the force-extension
relation at equilibrium. The novel description of the kinetics and energetics of the B-S transition of
ds-DNA improves our understanding the biological role of the S state in the interplay between
mechanics and enzymology of the DNA-protein machinery.
\end{abstract}

\maketitle

\section{Introduction}
Precise characterization of the mechanical properties of double stranded DNA (ds-DNA) is
fundamental for understanding the mechanism of the molecular machines used by cells to duplicate
and repair their genome and modulate the accessibility of the genetic information (1, 2). In living
cells, or in physiological solution in the absence of stress, the basic conformation of ds-DNA is the
B-form, that is with the two strands hydrogen-bonded to form a right-handed double helix, with a
periodicity of $\sim$ nm or, with a base-pair (bp) separation of 0.33 nm, 10.5 bp. Single bio-molecule
manipulation has allowed to investigate ds-DNA mechanics in the range of forces from a fraction of
piconewton to over 100 pN, that is in the force range exerted by proteins during their interactions
with DNA. At forces smaller than $\sim$10 pN the DNA backbone bends under the influence of random
thermal fluctuations (3-8); at forces above $\sim$10 pN, the molecule exhibits the stress-strain relation of
an elastic material, until the force attains $\sim$65 pN, when, within a few piconewtons of force change,
it modifies its molecular structure to a new ``overstretched'' form (S-form), that is characterized by a
length 70\% larger than that of the B-form (9, 10).

Control of the twist of the ds-DNA molecule (11, 12) has shown that its length is extremely
sensitive to the degree of twisting, revealing the role of the torque in the structural transition and
that the extended S-form of ds-DNA is characterized by an increase in bp per turn from 10.5 to 33-
37 (13, 14). Thus, untwisting of ds-DNA accompanies the overstretching transition and the torque
defines the force at which the transition occurs, according to a global phase diagram of the various
conformations of ds-DNA (13, 14). Within the force and torque ranges imposed by length or load
perturbations on a molecule that has a nick (a break in the covalent bonds of either of the two sugar-phosphate
backbones) or is free to rotate, the phase diagram at equilibrium is limited to the
definition of the boundary between the two forms B and S (blue line in Fig. 1a, reprinted with
permission from Sarkar et al. 2001 paper (13)). The thickness of the line, with a width of ~0.5 $k_BT$
in torque and ~4 pN in force, indicates that the transition is sharp, consistent with a cooperative
mechanism. The B-S boundary region is expanded in Fig. 1b, where each blue line indicates the
locus of force-torque points for which both states are contributing to the length of the molecule in
the proportion indicated by the numbers close to the line, while the open areas indicate the region of
the force-torque plane where the length fraction contributed by either state is $>$ 90\%.

\begin{figure}[h!]
\includegraphics[clip,width=8 cm]{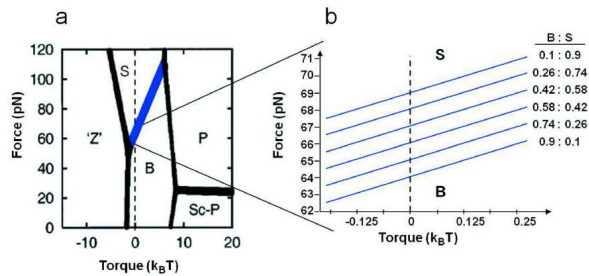}
\caption{The force-torque phase diagram of ds-DNA reprinted with permission from Sarkar et al. (13)
(a) The regions separated by diffuse borders represent the force-torque points for which the DNA
state assumes a pure structural phase. Inside the thickness of the borders, adjacent phases coexist.
Phases are named according to Sarkar et al. (13). For the purpose of this work the relevant phases
are the B-form, the basic conformation of ds-DNA, and the 1.7 times longer S-form, separated by a
thick blue line. The dashed vertical line represents the zero torque condition observed in our
experiments. (b) The thick blue line separating the S and B forms in (a) is expanded to include a
series of parallel thin blue lines with vertical distance 1 pN, each one representing the locus of
force-torque points with a given proportion (indicated by the numbers) of the contribution of the
two states to the length of the molecule.}
\label{1}
\end{figure}

In contrast to the above analysis, the double-stranded structure of the S form of DNA has been
challenged by previous work where it is suggested that the overstretching transition is due to force
induced melting and separation of the two strands (15-17). Evidences in support of this view are the
dependence of the force for the overstretching transition on the solution conditions such as
temperature, pH and ionic strength, as well as the presence of hysteresis in the rewinding to the B form.
Melting with strand separation during the overstretching transition has been recently recorded
with fluorescence microscopy (18, 19). However the interchange between a base-paired B state and
a single stranded S state cannot explain the overstretching transition when, as it occurs in the range
of temperatures below 35 °C, (i) hysteresis is asymmetric (upon release the first part of the force-extension
curve in the transition region coincides with the force extension curve upon stretch) (9,
10, 20) and (ii) the elastic modulus of the overstretched form is higher than both that of the B-form
and that of the single stranded DNA at the same force (21, 22). Recently it has been shown that
even ds-DNA that is topologically closed but rotationally unconstrained can undergo the
overstretching transition without hysteresis upon relaxation (23). Moreover, it has been shown that,
depending on salt concentration, temperature and DNA sequence, overstretching under force clamp
may involve either a rapid non-hysteretic transition to the elongated double stranded S form or a
slow hysteretic strand separation (24) and that, even in presence of nicks and free ends a stable
extended S-form of DNA can be generated in physiological conditions (25).

Until now, however, the transition kinetics between compact and extended states of ds-DNA has
not been measured. Moreover the existing estimates of the degree of cooperativity of the transition
vary by an order of magnitude (26, 27), the structural and energetic aspects of the B-S transition
have not yet been characterised and even the existence of the S state itself awaits direct
experimental evidence (28, 29). Here we describe the force-extension relation of the ds-DNA of 
$\lambda$-phage in physiological solution (150 mM NaCl, 10 mM Tris, 1 mM EDTA, pH 8.0 at 25 $^{\circ}$C), using
a double laser optical tweezers (10) that can impose on the molecule length changes made by a
staircase of force steps each complete within a few milliseconds and of size 0.5-2 pN via beads
attached to opposite strands. Under these conditions a step stretch applied in the region of forces of
the overstretching transition generates an elastic distortion on the molecule, consisting of a transient
negative internal torque that induces in the subsequent few hundreds of milliseconds an isotonic
elongation with exponential time course that implies untwisting towards a new equilibrium where
the length fraction contributed by the S state is increased at the expenses of that contributed by the
B state (Fig. 1b). Several steps are necessary for the elongation response to complete the
overstretching transition and the kinetics of the elongation response is explained by a simple two 
state model in which the DNA molecule is assumed to consist of a series of independent units of
$\sim$25 bps that can each flip between two states. The mechanism proposed also reproduces the
experimental equilibrium force-extension curves, whose lack of hysteresis shows that the extended
S form can be produced in a perfectly reversible way.

\section{Materials and methods}
{\bf \small Preparation of DNA.} A linearized double strand $\lambda$-DNA (48.5 kbp, New England Biolabs) has
been used for the experiments. Single molecules of DNA have been tethered between two
streptavidin-coated beads (Spherotech, Libertyville, IL) of either 3.28 $\mu$m-diameter (17 molecules)
or 2.18 $\mu$m-diameter (4 molecules). For labeling ds-DNA at both ends but on opposite strands, bio-
11-dCTP, dATP, dGTP, and dUTP were polymerized on opposite $\lambda$'s 12-bp sticky ends using
Klenow enzyme. The experiments were made at 25 $^{\circ}$C in solution with the following composition:
150 mM NaCl, 10 mM Tris, 1nM EDTA, pH 8.0.\newline

{\bf \small Optical tweezers.} Dual-beam optical tweezers (10, 30) were used to control both extension and
tension of single DNA molecules . In the sample chamber, a streptavidin-coated bead was held by
an optical trap and the second bead was positioned at the tip of a micropipette through suction (10).
The micropipette was fixed to the chamber, which in turn was mounted on a piezoelectric flexure
stage (PDQ375, Mad City Lab, Madison, WI), that controls the position of the pipette with an error
$<$ 1 nm. A diluted solution of DNA with biotinilated 3' ends was run through the cell. Once one end
of DNA molecule was attached to the trapped bead, the bead on the pipette was moved toward the
trapped bead until the opposite end of the molecule was bound. The procedure is identical to that
described by Bennink et al. (31). The extension of the molecule is measured by the movements of
the two beads (30). The force is measured by the position of the trapped bead, calibrated by means
of viscous drag. A change in force produces a movement of the bead that is measured by the change
in the light momentum with a precision of $\sim$0.2 pN (30). The absolute extension of the molecule is
estimated at the end of the experiment by moving the pipette toward the trapped bead and
measuring the position of the pipette, with 5 nm precision, when the two beads start to separate as
judged by the force signal.\newline

{\bf \small Force clamp experiment.} Our system operates either in length or in force clamp mode. In the first
case the feedback signal for the servo system that controls the position of the piezoelectric stage via
a proportional, integrative, and differential amplifier (PID) is the position sensor of the piezo itself,
while in the second the feedback signal for the servo system is the recorded force and the
piezoelectric stage moves to compensate for any difference with the desired force. When force is
clamped at a constant value, its standard deviation is  $\leq$ 0.3 pN. Force and extension of the molecule
are acquired at a rate of 1 kHz. For the experiment, the DNA molecule is first stretched in length
clamp mode at constant pulling rate from the starting unstressed length, until a pre-set force $F_\text{max}$,
above the overstretching force, is reached ($x_1$ in Fig. 2a). The system is then switched from length
clamp mode to force clamp mode and a staircase of force steps (size, $\Delta F$, 0.5-2 pN, interval 5 s) is
imposed on the molecule, first in release (negative staircase) down to a force $F_\text{min}$ below the force
for the overstretching transition ($x_2$ in Fig. 2a) and then in stretch (positive staircase) up to the preset
force $F_{\mathrm max}$ (Fig. 2a, blue trace). At this point ($x_3$ in Fig. 2a) the system is switched back to the
length clamp mode and the molecule is released at constant rate to the initial length. The setting of
the PID amplifier controlling the feedback gain is adjusted to minimise the error between the
stepwise command signal and the actual force step imposed on the molecule. During the
overstretching transition the complex compliance of the molecule (the sum of the changes in length
during and after the force step) becomes quite large and the gain of the proportional amplifier, set
for the stiff region of the molecular response, becomes insufficient to preserve a stepwise change in
force. On the other hand presetting from the beginning the feedback gain to the value necessary for
the compliant region of the molecular response would cause the gain to be too high for the initial
stiff region, which would lead the system to oscillate. The best possible shape of each step of the
positive staircase is attained by evaluating the amplitude of the length response and the
deterioration of the force step for each force level during the preceding negative staircase and
adjusting consequently the proportional gain of the corresponding step in the subsequent positive
staircase. For this reason the kinetic analysis is based only on the length responses to step increases
in force. The performance of the system provides that (i) in the region of the response below and
above the overstretching transition, the rise-time of the force step is $\sim$2 ms; (ii) in the region of the
overstretching transition the step rise-time increases, but even for the responses with the maximum
complex compliance it remains $\leq$20 ms, that is one order of magnitude shorter than the time
constant of the corresponding length response. An experiment is excluded from the analysis if even
only one element of the force staircase deteriorates, loosing the stepwise shape.
The size of the force step $\Delta F$ was limited to the range 0.5-2 pN, because for $\Delta F <$ 0.5 pN the signal
to noise ratio was too low and for $\Delta F >$ 2 pN the number of intermediate transitions was too small.
The analysis refers only to data from the 2 pN step (17 molecules) and 0.5 pN step (4 molecules) to
maximise the possible effects of the size of the step and thus of the extent of the elongation
response on the kinetics of the overstretching transition.

\begin{figure}[h!]
\includegraphics[clip,width=8.0cm]{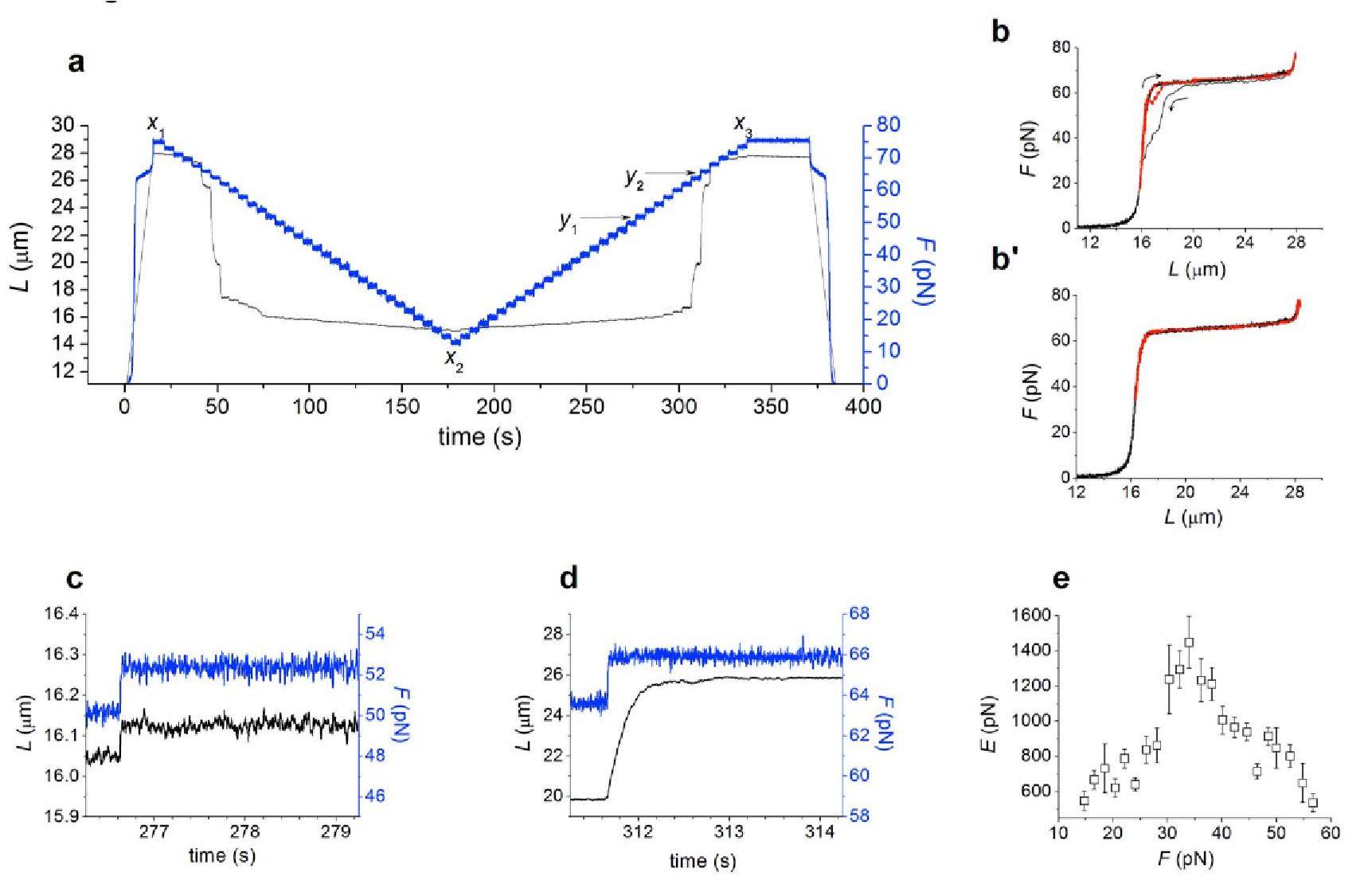}
\caption{Response to stretch-release of ds-DNA in length clamp (ramp shaped length change) and in
force clamp (staircase of 2 pN force steps at 5 s interval). (a) Change in length (black line) and force
(blue line) during the protocol used for the experiment. At zero time the molecule is stretched with a
ramp in length clamp mode up to a force (76 pN) beyond the region of the overstretching transition,
that is made evident by the transient reduction in slope of the force-time trace; at $x_1$ the feedback is
switched to force and the molecule is released with the negative staircase of force steps down to 14
pN ($x_2$) and then stretched with the positive staircase of force steps to attain again 76 pN ($x_3$); then
the feedback is switched back to length and the molecule is released down to zero force with a
ramp. (b) Force-extension curves in length clamp (black) and in force clamp (red) drawn from a
protocol like that in Fig. 2a for either a molecule that shows hysteresis in relaxation (b) or a
molecule without hysteresis (b). In (b) the direction of the length change along the traces is
indicated by the closest arrow. (c) Expanded record of the length response (black) to the force step
(blue) 50-52 pN (corresponding to $y_1$ in Fig. 2a). The length response is purely elastic. (d)
Expanded record of the length response (black) to the force step (blue) 64-66 pN (corresponding to
$y_2$ in Fig. 2a). The elastic length change is followed by a slower and much larger elongation with an
exponential time course. (e) Change in elastic modulus ($E$) with force measured by means of force
steps in the range (15-60 pN) where the molecule in B-state exhibits an intrinsic elasticity. Data are
mean $\pm$ SD from 14 molecules with 2 pN step.}
\label{2}
\end{figure}

\section{Results and discussion}
{\bf \small Stretching the DNA molecule by a staircase of force steps.}
In Fig 2a the molecule, tethered
between two streptavidin coated beads by means of its biotinilated 3' ends, is first pre-stretched in
length clamp mode at constant pulling rate to a length ($L = 28 \mu$m, $x_1$ in Fig. 2a) at which the force
($F$ = 76 pN) is above the overstretching transition, then the control is switched to force clamp mode
and the molecule is released with a staircase of force steps of 2 pN and 5 s interval down to 14 pN. 
The molecule is then stretched with the same force step staircase up to 76 pN, where the
control is switched back to the length clamp mode and the molecule is released at a constant rate to
its unstressed length. During the negative staircase ($x_1 \rightarrow x_2$) the molecule starts to respond
according to its intrinsic elasticity in state S and then crosses the region of the overstretching
transition, that includes five-eight force steps. Conversely during the positive staircase ($x_2 \rightarrow x_3$) 
the
molecule starts in state B and, once attained the threshold for the overstretching transition, takes
five-eight force steps to acquire the 100\% S state.
The length response to the force step (rise-time $t_r\sim$2 ms) in the region of the intrinsic elasticity in
state B ($y1$ in Fig. 2a, $F$ = 50 pN) is shown at higher gain in Fig. 2c. The 2 pN force step induces a
stepwise increase in length of $\sim$60 nm, corresponding to a compliance of $\sim$30 nm pN$^{-1}$. The length
response to the same force step in the region of the overstretching transition ($y_2$ in Fig. 2a, $F$ = 66
pN) is shown in Fig. 2d at ten times lower gain than that in Fig. 2c. In this case the initial elastic
response is followed by a roughly exponential isotonic lengthening of amplitude $\sim$5 $\mu$m, complete
within the first 500 ms of the 5 s interval.
The force-extension relation obtained from this experiment (Fig. 2b) shows that, at the resolution
necessary to describe the whole process, the differences between the two protocols, constant pulling
rate (black line) and force step staircase (red line), emerge only during the relaxation phase: when
relaxation is induced by a negative staircase of force steps, the hysteresis in the structural transition
is reduced. This could be due to the fact that the relaxation obtained with the negative staircase is
preceded by a ramp-shaped lengthening that takes a shorter time to stretch the molecule, compared
to the time taken by the positive staircase that precedes the length clamped release. In 4 of the 17
molecules used for the 2 pN force step there was no sign of hysteresis upon release (Fig. 2b) either
in length clamp (black trace) or in force clamp (red trace). The chosen mechanical protocol, pulling
from different strands, in principle allows strand separation, thus the lack of such rupture events for
molecules overstretched up to 76 pN, shows that the DNA remains double-stranded as it unwinds to
achieve the extended conformation and therefore that melting and generation of conformations
more stable than the S-state, such as single stranded (ss) DNA (15-19), is not a necessary condition
for the overstretching transition.
In the region of forces $<$ 60 pN, where the B state of ds-DNA exhibits an intrinsic elasticity, the
elastic modulus E can be calculated from the records as in Fig. 2c. E shows a biphasic dependence
on the force at which the step is applied (Fig. 2e): in the range of forces 15-35 pN E increases from
550 $\pm$ 50 pN (mean and SD, 14 molecules with 2 pN steps) to a maximum of 1450 $\pm$ 130 pN (10,
22, 32) and then, in the range of forces 35-57 pN, it reduces again to 530 $\pm$ 60 pN. This switching
between two kinds of elasticities fits well with the finding that, at forces $<$ 35 pN, the twist-stretch
coupling of the B state of DNA is negative, that is the molecule overwinds under stretch and
reverses only at higher forces (32).\newline

{\bf \small Load dependence of the extent and rate of the ds-DNA elongation.} 
Above 60 pN five-eight
steps of 2 pN are necessary to complete the B-S transition (Fig. 2a). The number of steps increases
when the step size is reduced, so that there is an inverse proportionality between number and size of
the force steps necessary to complete the transition. The elongations elicited by a series of six 2 pN
steps, identified by the color and force attained at the end of each step, are shown in Fig. 3a. Going
from the first (62 pN, purple) to the sixth (72 pN, green) step, the amplitude of the elongation ($\Delta L_e$)
increases abruptly up to a maximum of $\sim$3.3 $\mu$m (third step, 66 pN, black) and then reduces again.
The initial speed of the elongation does not vary in proportion to its size, so that the time for
completing the elongation is shorter for the first and sixth step and increases with the size of the
elongation. After the elastic response has been subtracted, the time course of the elongation ($\Delta L$) in
response to a step increase in force (Fig. 3a) can be interpolated by the exponential equation $\Delta L =
\Delta L_e \left(1-\exp(-rt)\right)$, where $t$ is the time elapsed after the step, $r$ is the rate constant and $\Delta L_e$  is the
asymptotic value of the elongation. $r$ is larger at the beginning and at the end of the B-S transition.
The dependence of $r$ on the force attained after the 2 pN step is shown for 17 molecules in Fig 3b
(blue symbols). It can be seen that $r$ points from the 4 molecules without hysteresis on release
(squares) superposes on those from the 13 molecules with hysteresis (triangles). The $r$-$F$ relation is
U shaped, showing maxima of $\sim$50 s$^{-1}$ at the beginning ($F \sim$62 pN) and at the end ($F \sim$72 pN) of the
overstretching transition and a minimum of 3.82 $\pm$ 0.68 s$^{-1}$ (mean and SD, calculated in the range
65.5 - 66.5 pN) in the plateau region. The relation obtained from 4 molecules for which the size of
the step in the staircase was 0.5 pN (green circles) almost superposes on that with 2 pN steps (r in
the range 65.5-66.5 pN is 3.94 $\pm$ 1.52 s$^{-1}$, the same as that obtained with the 2 pN step), indicating
that $r$ is independent of the force step size and depends only on the final force attained during the
step. The monotonic increase of the total length of the molecule $L$ with the increase in force (Fig.
2b) implies that also the $r$-$L$ relation is U shaped (see Fig. S2 in SI).
The findings that the rate of elongation has the same U shaped relation versus either the force (Fig.
3b) or the length of the molecule (Fig. S2), independently of the size of the force step, are direct
indications that the response is an intrinsic property of the molecule, related to its transition
kinetics, and does not depend on the viscosity of the medium. Further evidence, in relation to either
the translational drag of the bead or the rotational drag of the molecule while untwisting, is given in
detail in SI 1.\newline

\begin{figure}[h!]
\includegraphics[clip,width=8 cm]{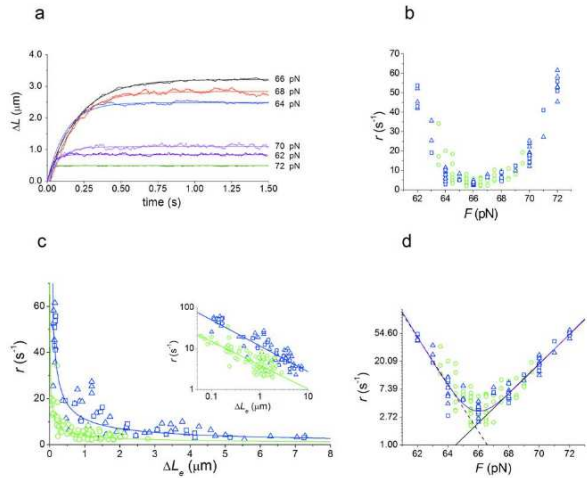}
\caption{Dependence of the extent and rate of elongation on force. (a) Superimposed time courses of 
elongation ($\Delta L$) following the series of 2 pN force steps starting at 60 pN during the staircase. The
level of force attained by the step is reported next to the trace. The lines are the exponential fits to
the traces as identified by the force level. The rate constants are 47 s$^{-1}$ (62 pN, purple), 6.1 s$^{-1}$ (64
pN, blue), 3.8 s$^{-1}$ (66 pN, black), 5.1 s$^{-1}$ (68 pN, red), 9.2 s$^{-1}$ (70 pN, pink) and 58 s$^{-1}$ (72 pN, green).
(b) Relationship between the rate of elongation (r) and force (F) following 0.5 pN (green circles)
and 2 pN (blue symbols) force steps. 2 pN data pooled from 13 molecules showing hysteresis in
relaxation (triangles) and 4 molecules without hysteresis (squares). (c) Relationship between $r$ and
the final length change induced by a step $\Delta L_e$ for the same data (and symbols) as in (b). The lines
are the fit on either 2 pN data (blue) or 0.5 pN data (green) by the power equation $r = C \Delta L_e$
b. The
inset shows the same data plotted on a double logarithmic scale. The lines are the linear regression
fit to data: the regression coefficient, b, is -0.61 ± 0.04 for the blue line (2 pN) and -0.55 $\pm$ 0.05 for
the green line (0.5 pN). b corresponds to the exponent of the power equation in the plot with linear
scale. (d) Relation between $\ln r$ and force for the same data as in Fig. 3b. The black straight lines
are the fits of Eq. 3 (expressing $k_+$, continuous line) to pooled 2 pN and 0.5 pN data in the force
range 67-72 pN and Eq. 4 (expressing $k_-$ , dashed line) to data in the range 61-65 pN. The purple
line is the four parameter fit (see SI 3, Eq. S8) to pooled 2 pN and 0.5 pN data.}
\label{3}
\end{figure}

The amplitude of the elongation ($\Delta L_e$) following the force step varies with force in a way that is a
mirror image of the rate. The maximum elongation occurs in the central region of the overstretching
transition: in the range 65.5 -- 66.5 pN, where $r$ is minimum, $\Delta L_e$  is 5.03 $\pm$ 1.63 $\mu$m for the 2 pN step
and 1.34 $\pm$ 0.51 $\mu$m for the 0.5 pN step. The relationship between $r$ and $\Delta L_e$  appears hyperbolic
(Fig. 3c) with a larger curvature for the smaller force step. A log-log plot of this relation shows that,
whatever the force step size is, $r$ is related to $\Delta L_e$  through the power equation: $r = C  \Delta L_e b$ , where b
is $\sim$-0.6 (the regression coefficient $b$ is -0.61 ± 0.04 and -0.55 $\pm$ 0.05 for the 2 pN and 0.5 pN step
respectively). In this case it can be demonstrated (see SI 2) that the elongation following a force
step in the region of the overstretching transition is consistent with a two-state reaction with first
order kinetics. Moreover, the downward shift observed for the smaller force step (corresponding to
a reduction of $r$ by 3.5 times for $\Delta L_e$  = 1 $\mu$m) indicates once again that the rate of elongation is not
determined by the extent of elongation, a result that confirms that the rotational drag of the
molecule while untwisting is not the relevant parameter for the kinetics of the process (see SI 1B).\newline

{\bf \small The structure of the transition state.}
Assuming the DNA molecule as an ensemble of independent
units of uniform length that exist in either a compact (B) or an extended (S) state, the equilibrium
distribution at each force is defined by the free energy difference between the two states and the
work done to promote the transition. In this case the rate constants for the elongation ($k_+$) and the
shortening ($k_-$) of the molecule depend exponentially on the force according to Kramers-Bell theory
\begin{equation}
\label{kramI}
k_+=A_+ \exp\left(F x^\ddagger_+ \over k_B T \right)
\end{equation}
\begin{equation}
\label{kramII}
k_-=A_- \exp\left(F x^\ddagger_- \over k_B T \right)
\end{equation}
where $x^\ddagger_+$ and $x^\ddagger_-$ (nm) are the distances to the transition state from the compact and the extended
states respectively, $k_B$ is the Boltzmann constant, $T$ is the temperature in Kelvin and $A_+$ and $A_-$ (s$^{-1}$) are
the rate constants at zero force.

The unidirectional rate constants expressed in logarithmic units have a linear dependence on force
and the slope of the relation is the distance from the starting state to the transition state divided by
$k_B T$:

\begin{equation}
\label{kramIII}
\ln k_+=\ln A_+ \left(F x^\ddagger_+ \over k_B T \right)
\end{equation}
\begin{equation}
\label{kramIV}
\ln k_-=\ln A_- \left(F x^\ddagger_- \over k_B T \right)
\end{equation}

In our experiments the observed rate constant $r$ (Fig. 3b) is the sum of the forward and reverse rate
constants. However, since the equilibrium is dominated by $k_-$ in the low force side of the
overstretching transition and by $k_+$ in the high force side, we used the logarithmic relations in these
two regions to separately estimate the slopes of the respective unidirectional reactions and thus of
the distances to the transition state (Fig. 3d). The parameters of the linear regression equations fitted
to ln $r$ with Eq. 3 in the force range 67-72 pN and Eq. 4 in the range 61-65 pN are reported in Table
1 either for the 0.5 pN steps (green circles) or for the 2 pN steps (blue symbols). It can be seen that
in either case all the kinetic parameters are not significantly influenced by the force step size (P>
0.1). The linear fits on the pooled 0.5 pN and 2 pN data within the same force ranges give similar
values: (i) $\ln A_+$ = -34.14 ± 1.11 and $x^\ddagger_+$= 2.18 ± 0.07 nm for Eq. 3 (black continuous line), (ii) 
$\ln A_+$= 56.05 ± 2.99, $x^\ddagger_-$= 3.67 ± 0.19 nm for Eq. 4 (black dashed line).

\begin{table}[ht]
\begin{tabular}{|c|c|c|c|c|}
\hline
 \hfil &   \multicolumn{2}{|c|}{Eq. 3} & \multicolumn{2}{|c|}{Eq. 4}  \\
\hline
Force step (pN)    & ln $A_+$  & $x^\ddagger_+$ (nm) & ln $A_+$   & $x^\ddagger_-$ (nm) \\
\hline
0.5    & -32    ± 7   & 2.1  ± 0.4  & 64 ± 12 & 3.9 ± 0.8 \\
2      & -34.5  ± 1.3 & 2.18 ± 0.07 & 61 ± 3  & 3.7 ± 0.2 \\
pooled & -34.1  ± 1.1 & 2.18 ± 0.07 & 56 ± 3  & 3.7 ± 0.2 \\
\hline
\end{tabular}
\caption{Kinetic parameters of the two state reaction estimated with Eq. 3 and Eq. 4.}
\label{tabone}
\end{table}

Alternatively, a more complex, four-parameter fit with the expression of the relaxation rate $r = k_+ +
k_-$ (See SI 3, Eq S8) can be performed on the entire force range (purple line in Fig 3d) and gives
values of the four parameters $A_+$, $A_-$, $x^\ddagger_+$ and $x^\ddagger_-$ that are almost identical to those determined here by
fitting separately the low and high force branch of the relation.
From the estimated distances to the transition state, $x^\ddagger_+$ and $x^\ddagger_-$, we conclude that the transition
occurs roughly midway between the B state and the S state of DNA, slightly shifted toward the B
state. The sum of the two distances, ($x^\ddagger_+ + x^\ddagger_-$), gives a total length change for the unitary reaction
$\Delta x$ of 5.85 $\pm$ 0.20 nm. The ratio of $\Delta x$ over the elongation undergone by each bp (0.33*1.7 - 0.33 =
0.231 nm) measures the number of bp involved in the unitary reaction, or cooperativity coefficient,
and is 25.32 $\pm$ 0.86.
A test is conducted in SI 4 on the degree of variability of the size of the unitary reaction that is still
compatible with the observed equilibrium kinetics (the force-extension relation) and relaxation
kinetics (the force dependence of the rate of elongation following a force step). We demonstrate that
a stochastic variability up to 30\% in either $x^\ddagger_+$ or $x^\ddagger_-$ does not substantially affect either of the kinetic
parameters. Instead, the variability of either the free-energy barrier $\Delta G^\ddagger_+$
+ or the free energy change $\Delta G$ must be lower than 10\%.

Apparently the reaction kinetics under our experimental conditions is not affected by the fact that
the return from the extended to the compact state exhibits or not hysteresis (compare blue triangles
and squares in Fig 3b-d), indicating that the presence of nicks, revealed in some molecules by the
hysteretic behaviour, does not affect the kinetics of the B-S transition. Considering that the
occurrence of nicks reduces the length of the units rotating as a whole, this finding shows that the
rate of elongation does not depend on the length of the rotating unit and thus further supports the
view that the elongation kinetics is not affected by the rotational drag (see SI 1B).
The finding that the transition state is almost midway between the compact and extended states
implies that the force (Fm) at which the relaxation rate is minimum also corresponds, within our
resolution limit, to the force (Fe) at which the work done for the elongation (We) equals the free
energy difference between the compact and the extended state: $W_e = F_e  \Delta x$ (see SI 2). With $\Delta x$ =
5.85 nm, $W_e$ is (66 * 5.85 =) 386 zJ per molecule (or 236 kJ per mol), that at 25 $^{\circ}$C corresponds to
94 $k_BT$. The average binding free energy per bp obtained from these mechanical measurements is
($F_e$ * 0.231 nm =) 15.25 zJ, corresponding to only 3.71 $k_BT$ per molecule. The cooperative
mechanism for DNA elongation increases the stability of the DNA structure in the B state, by
increasing the minimal free energy change necessary for the elongation reaction to 94 $k_BT$.\newline

{\bf \small Comparison with recent experiments.}
The exponential kinetics observed for $\lambda$-DNA apparently contrasts with recent observations of
similar overstretching experiments under force clamp. Fu and coworkers (25) report on constant
force DNA overstretching experiments performed by means of magnetic tweezers. Under melting preventing
conditions (high salt concentration or moderately GC rich sequences) DNA constructs of
approximately 600 bp show, in response to force increase, lengthening followed by length
fluctuations, instead of a smooth exponential elongation. In order to assess the predictions of our
two-state model for such molecular size we performed a series of Monte Carlo simulations. A set of
23 two state units, each 25 bp long, was defined in order to simulate a molecule 575 bp long. All
units were initially assigned a compact conformation and force was gradually increased with 1 pN
steps. For each force value the system was integrated for 20 seconds according to a stochastic
dynamics where each unit could change its state from compact to extended and from extended to
compact according to their transition probabilities computed as $k_\pm dt$ with dt the integration time. 
$k_\pm$ were computed according to Eq. 1 and 2 using the parameters resulting from the fit of the rate-force
relation as in Fig. 3a, b. Fig. 4a confirms that, for a molecular size as low as 575 bp, random length
fluctuations are so prevalent to mask almost completely the shape of the relaxation. On the contrary,
as shown in Fig 4b, that reports the result of a simulation with 1936 two state units, or (1936 · 25 =)
48,400 bp, the entire DNA molecule exhibits a much smoother lengthening, that with an adequately
expanded time scale, reveals its exponential kinetics. This is due to the self averaging effect of
length fluctuations of the units in a linear chain, when the number of units is sufficiently high: for
each extending unit another one contracts compensating the effect of the first. A striking result of
this simulation is that the number of bp involved in the unitary reaction, selected for fitting the
responses of the entire molecule with the two state model, is able to predict with a very good
approximation the length fluctuations found by Fu and coworkers with the shorter molecule.\newline

\begin{figure}[h!]
\includegraphics[clip,width=6.5cm]{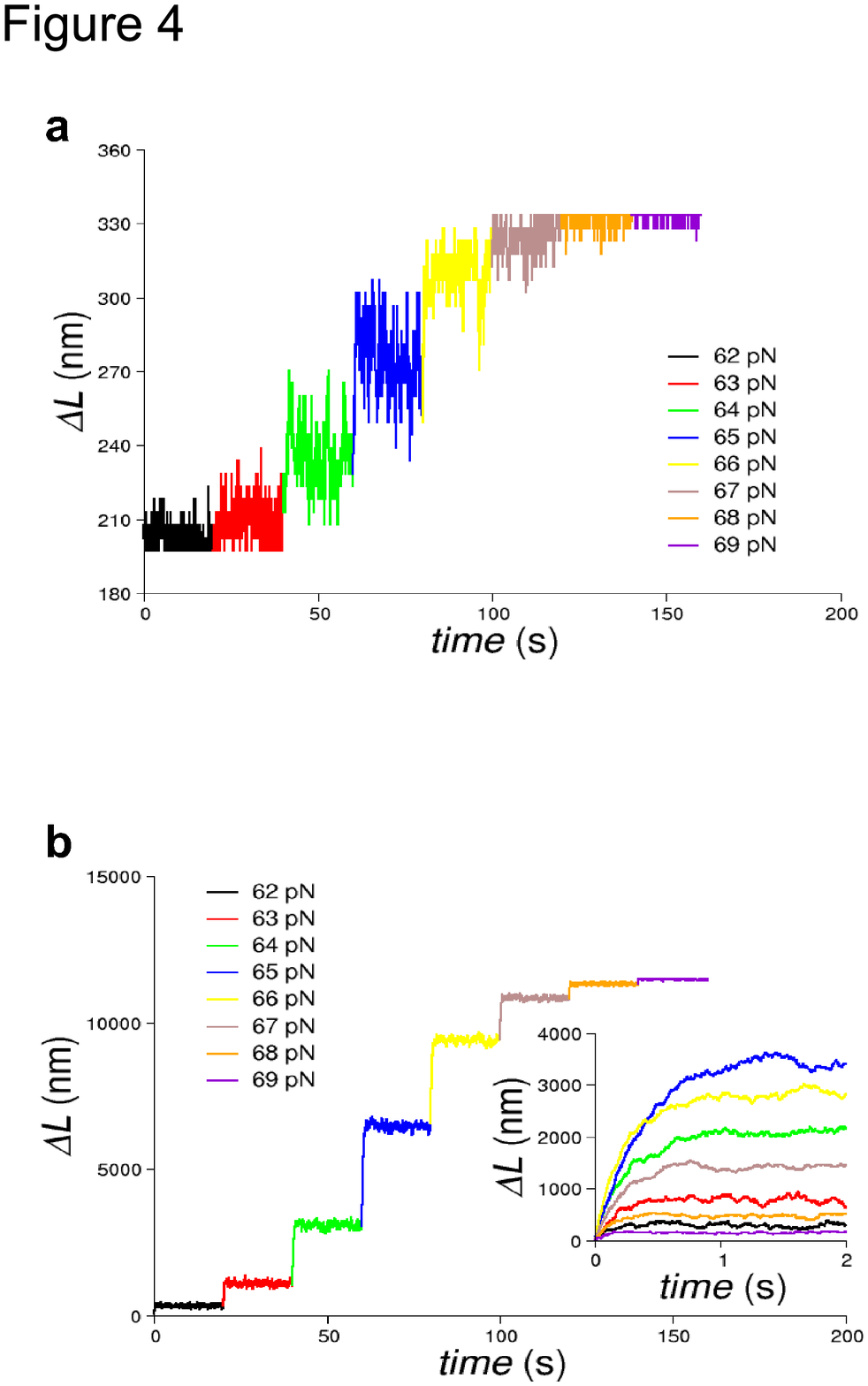}
\caption{Simulated time course of eight consecutive length changes in response to a staircase 1 pN
force steps. The responses are calculated for a system composed of identical two-state units of 25
bp. (a) 23 two state units (575 bp); (b) 1936 two state units (48,400 bp). Different colors correspond
to different force values from 62 to 69 pN. The inset in (b) shows the superposed length responses
on a faster time base, adequate to resolve the exponential time course.}
\label{4}
\end{figure}

{\bf \small The extended state of ds-DNA.}
The results of this work give solid basis to the definition of the S
state of ds-DNA, showing that 100
overstretching transition remains double-stranded and is produced in a perfectly reversible way via
intermediate state transitions. This conclusion agrees with the recent evidence that also
topologically closed ds-DNA can undergo the overstretching transition at $\sim$65 pN (23). In a
molecule free to rotate or unpeel, further transitions to the more stable ss-DNA is not directly
responsible for the overstretching transition and would take place either on a larger time scale or
under the action of melting promoting factors such as temperature, pH and ionic strength (16, 17,
21, 22, 25, 33). Thus the strand separation during the overstretching transition recently
demonstrated with fluorescence microscopy (18, 19), was likely promoted at some stage of the
specific molecule manipulation.
The presence of asymmetry in the hysteresis of the force-extension curve as well as the high rigidity
(22) and helicity (9, 10, 12, 14) of the extended DNA state can be reproduced using a four-state
model that includes a base-paired B state, a base-paired S state, a molten state and a unpeeled state
(21). However, the first description in the present work of the relaxation kinetics of intermediate
states of the overstretching transition (Fig. 3) together with the demonstration that the force-extension
relation at equilibrium may be perfectly reversible (Fig. 2b) indicate that the whole interconversion
from the B state to the S state can be explained in terms of a two state reaction in which
the double-stranded structure is preserved. Under these conditions we can define the size of the
unitary reaction ($\Delta x \sim$5.8 nm) and provide some structural insight on the transition state.
The definition of the kinetics and energetics of the B-S transition of ds-DNA in this work represents
a significant contribution to understanding the biological role of the S state in the interplay between
mechanics and enzymology of DNA-protein complexes. DNA overstretching promotes the
nucleation/polymerisation on the DNA of the RecA protein (34), that is the prerequisite for
homologous recombination between two DNA segments (35). However the extended form of ds-
DNA is also likely to be involved in the regulation of other basic functions of the DNA-protein
machinery.
X-ray crystallography (36) has allowed the structural details of the boundary between the B state
and the Z state, the left-handed conformation associated with the transcription start site, to be
described (37, 38). As shown in Fig. 1a, the crystallographic approach is possible because the B-Z
transition occurs under no- or low-load conditions and for slightly negative torque values that are
likely induced (or stabilised) by the Z-DNA binding proteins (39). Instead, the boundary for the S-Z
transition lies in a region of higher loads and more negative torques. This implies that, even with the
molecule under stress, the torque contribution by the DNA-binding protein may not be adequate for
promoting the transition or, in other words, that in the S state the formation of the transcription start
site is inhibited, because in this state the DNA is more resistant than in the B state to the effect of
the negative torque. At the same time the position of the boundary for the S-Z transition in the
phase diagram of Fig. 1a makes it explicit that the S-Z transition can only be studied by integrating
time resolved structural and mechanical methods and not by X-ray crystallography, since the stress
on the DNA molecule cannot be reproduced in crystals. In this respect, the present description of
the load dependence of kinetics and energetic of the B-S transition is an important step forward in
understanding the role of the S state in gene regulation and expression.

\section*{FUNDING}
This work was supported by Ente Cassa di Risparmio di Firenze (Italy) and by Italian Institute of
Technology [SEED-MYOMAC 2009].

\section*{ACKNOWLEDGEMENTS}
We thank Malcolm Irving and Gabriella Piazzesi for comments on the manuscript.


\renewcommand{\theequation}{S.\arabic{equation}} 
\renewcommand{\thefigure}{S.\arabic{figure}}
\setcounter{equation}{0}  
\setcounter{figure}{0}

\section*{SUPPORTING INFORMATION}  

\subsection*{1. Influence of viscosity on the kinetics of the elongation-untwisting of the ds-DNA during the overstretching transition.}

\noindent
{\bf \small A. Drag produced on the trapped bead by the viscosity of the medium.}
The change in position
of the trapped bead during the movement of the piezoelectric stage reliably measures the tension on
the molecule only if it is not influenced by the drag due to the movement of the solution
accompanying the movement of the stage. The drag on the bead is $F_\eta=6 \pi \eta R V$ , where $\eta$ is the
viscosity of the solution (10-3 Pa s, at 25 $^{\circ}$C), $R$ is the radius of the bead (1.64 or 1.09 mm) and $V$ is
the translational velocity of the bead. Considering that the stiffness of the molecule is ~60 pN mm$^{-1}$
and the stiffness of the trap is 150 pN mm$^{-1}$, a step of 2 pN complete in 2 ms implies a bead
movement of ~40 nm at a velocity of ~20 mm s$^{-1}$. Consequently, for the bead with $R$ = 1.09 mm, $F_\eta$ 
attains a value of 0.5 pN and decays with a time constant of 0.5 ms (1/4 the rise-time of the step).
This analysis indicates that the viscous drag on the bead does not significantly influence the
position of the bead during the step nor the observed elongation kinetics. A direct test of this
conclusion is obtained by comparing the $r$-$F$ relations obtained with different bead diameters. In
Fig. S1a and S1b, the blue points data from in Fig. 3d and 3c (inset) respectively are unpooled to
identify those obtained with beads of 3.28 mm diameter (black symbols, 12 molecules) and 2.18 mm
diameter (red symbols, 5 molecules). In this way it becomes evident the absence of any effect of the
bead diameter on the overstretching kinetics. Eventually, it must be mentioned that the relaxation
rate expected in the case of a translational damping process solely depends on the bead size which
is constant during the entire experiment. It is therefore impossible that a convex profile of the $r$-$F$
relation with rates spanning more than one order of magnitude is determined by translational drag

\begin{figure}[]
\includegraphics[clip,width=8.5 cm]{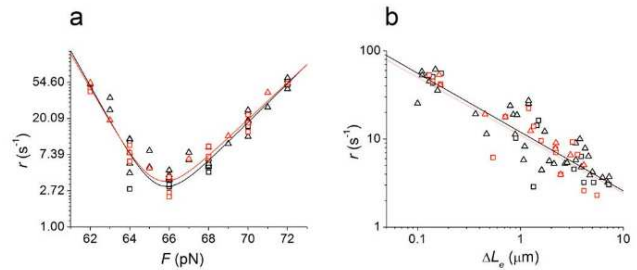}
\caption{(a): $r$-$F$ relation for the 2 pN steps from Fig. 3d blue symbols. (b): $r$-$\Delta L$ relation for the 2
pN steps from the inset of Fig. 3c blue symbols. Red symbols refer to data obtained with 2.18 mm
bead diameter (5 molecules); black symbols refer to data obtained with 3.28 mm bead diameter (12
molecules)}
\label{S1}
\end{figure}

\noindent
{\bf \small B. Rotational drag of the molecule while untwisting.}
A rotational drag of the ds-DNA while untwisting in response to a rise in torque has been directly
measured by attaching a bead near a nick and determining the angular velocity(1). In this way it has
been shown that the untwisting takes several minutes and the drag dominates the elongation-untwisting
velocity. However in that experiment the drag should be several time larger than in our
experiment, as it is generated by the revolutions of a large bead accompanying the untwisting of the
molecule.

During the overstretching transition under our conditions, the molecule of DNA elongates by 11 mm
while it reduces the number of turns from 4500 to 1450, that is it untwists by 278 turns mm$^{-1}$. The
largest elongation in response to a 2 pN step is ~5 mm attained within 0.5 s (Fig. 3). This implies a
rotational speed w of (278*5/0.5 =) 2780 turns s$^{-1}$, so that each of the two ends of the molecule
should counter-rotate at 1390 turns s$^{-1}$. According to measurements of rotational drag in the
experiment of Thomen et al. (2), where the two strands of DNA are attached to two independent
beads and separated at different velocities, a torque of 0.6 $k_BT$ would be necessary for the rotation at
the speed of our overstretching transition. However, in Thomen et al. experiment the rotating stretch
is at longitudinal force zero (and therefore the molecule is not straight), while during the
overstretching transition the longitudinal force is ~65 pN and the molecule can be assimilated to a
rigid rod. According to Levinthal \& Crane (3) the rigid rod model leads to a torque
$\tau=4 \pi \mu R_H L_\text{eff}$, where $\mu$ is the viscosity of the solution as above, $R_H$ the hydrodynamic radius
of DNA (1.05 nm (2)) and $L_\text{eff}$ the extension of the portion of DNA which rotates in order to
release the torsional stress. The molecular extension in the middle of the plateau of the
overstretching transition is approximately 22 mm which, under the assumption of torsional stress
accumulating in the middle of the molecule, gives a $L_\text{eff}$ of 11 mm. With $\omega$ = 1390 turns s$^{-1}$ 
(= 8730 rad s$^{-1}$) the maximal frictional torque results to be 0.3 $k_BT$. If, however, the torsional stress is
distributed uniformly along the whole length of the molecule, the resulting frictional torque should
drop to 0.15 $k_BT$. A frictional reaction of 0.15-0.3 $k_BT$ is expected to have a negligible effect on the
kinetics of the B-S transition as demonstrated below.

\begin{figure}[]
\includegraphics[clip,width=7.5cm]{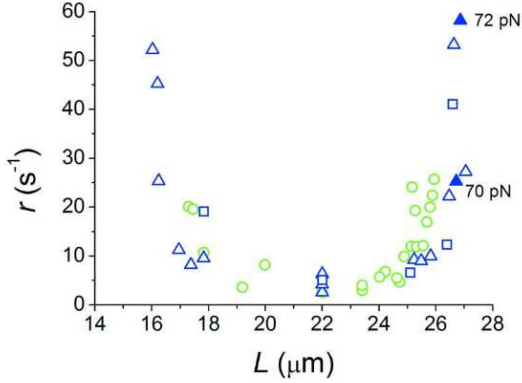}
\caption{Relation of the elongation rate ($r$) versus the molecular length ($L$) following 0.5 pN (green
circles) and 2 pN (blue symbols) force steps. 2 pN data pooled from 13 molecules showing hysteresis
in relaxation (squares) and 4 molecules without hysteresis (triangles). Data points are from the
same experiments as Fig 3b. Figures close to filled symbols indicate the respective forces.}
\label{S2}
\end{figure}

The presence of an external torque (in this case of frictional origin) opposing the B-S transition not
only modifies the free energy difference between S and B states but also the free-energy barriers
that the system must overcome for the transition (see Fig. S3). The B-S barrier will be increased
(transition more difficult) while the S-B barrier will be decreased (transition easier). Let us define $x_B$
and $x_S$ the distances between consecutive bps in the two states and \u03b8B and \u03b8S the twist angles between
consecutive bps, with numerical values: $x_B$ = 0.34 nm, $x_S$= 0.58 nm $\theta_B$=1/10 turns = 0.63
rad, $\theta_S$= 1/30 turns = 0.21 rad. Assuming for simplicity that the transition state is in the middle between
both $x_S$ and $x_B$ and $\theta_S$ and $\theta_B$, the change in barrier height due to the presence of a viscous
torque can be estimated and compared with the analogous change due to the presence of an external
force. The maximum torque induced barrier change is 
$\Delta E_\tau=\tau(\theta_S-\theta_B)/2=0.3 k_BT 0.4/2=0.25$ pN nm or 0.06 $k_BT$, 
while the barrier change caused by the external force $F$ (~65 pN at the transition)
is $\Delta E_F = F (x_S - x_B)/2$ = 65*0.24/2 = 7.8 pN nm or 1.9 $k_BT$. Thus, since the barrier change 
introduced by the viscous torque is smaller by a factor of 32 relative to the barrier change caused by the
external force, the analysis presented in the text, which disregards this contribution, is correct.
A simple direct test of the influence of the rotational drag of the molecule on the kinetics of
elongation is obtained by plotting the elongation rate $r$ as a function of the length of the molecule $L$
during the overstretching transition (Fig. S2). If the elongation kinetics was dominated by the
viscous friction one would expect the relaxation rates to depend on $L$. Actually the $r$-$L$ relation
shows a U shaped dependence that is hardly compatible with the hypothesis that the rotational drag
determines the elongation rate.

\begin{figure}[]
\includegraphics[clip,width=6.5cm]{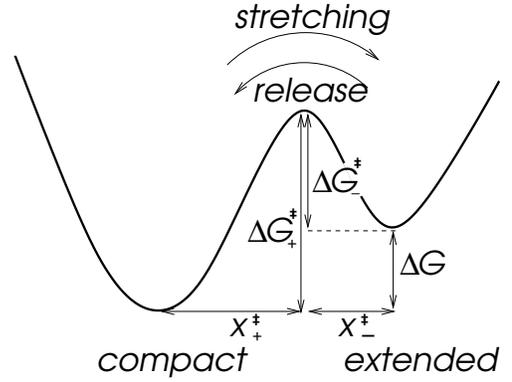}
\caption{Pictorial representation of the free energy landscape of a two-state unit. $\Delta G$, free energy
difference between the compact state and the extended state; $\Delta G^\ddagger_+$, energy barrier for the forward
transition; $\Delta G^\ddagger_-$, energy barrier for the backward transition. The distance of the transition state from
the compact state is $x^\ddagger_+$ and from the extended state is $x^\ddagger_-$.}
\label{S3}
\end{figure}

Another test is provided by the comparison of the rate of elongation in molecules that exhibit or not
hysteresis (blue squares and blue triangles respectively in Fig. 3b-d). Assuming that the presence of
nicks is revealed by the presence of hysteresis, it is expected that the length of independently
rotating units during the overstretching transition is shorter for blue squares than for blue triangles.
Consequently the finding that $r$ is the same for the same force supports the idea that the rotational
drag has not influence on the overstretching kinetics.
A further test is provided by the mechanical protocol described in the text, that allows to compare
the responses to 2 and 0.5 pN force steps. For the same force, the extent of elongation (and thus w)
is smaller with the smaller step (see text). Since the rotational drag depends linearly on $\omega$, if it was
dominating the kinetics of the process, it would have generated a larger rate constant of elongation
for the smaller step. However the observed rate constant is the same at the same force for either step-0.3 a
size (Figs. 3b and 3d, blue 2 pN, green 0.5 pN). On the other hand, when the rate constant of
elongation is plotted against the extent of elongation (Fig. 3c), the relation for 0.5 pN is shifted
downward with respect to that for 2 pN, as predicted by the two state model (see Eq. S7 in SI 2).
Both the theoretical treatment and the experimental evidences given above support the conclusion
that the rate of DNA elongation following force steps applied in the overstretching transition region
is mainly determined by the kinetics of the two state reaction. The conclusion is supported by the
findings described in the text that: (i) $r$ depends only on the force at the end of the step (Fig. 3b, d)
and does not depends on neither the step size nor the length of the molecule, (ii) the slope of 
the $\log r-\log \Delta L_e$ relation (Fig. 3c inset) is -0.6, $\ll$ 1, (see SI 2)

\subsection*{2. Assessing the validity of the two state reaction assumption for the overstretching transition.}

Here we verify that the kinetic model proposed is able to account both for the equilibrium and the
relaxation kinetics of the overstretching transition. According to a two state model, ds-DNA is
composed of an ensemble of units which can attain two different conformational states, a compact
B state, characterized by a molecular extension of 0.33 nm per bp, and an extended S state,
characterized by a molecular extension of 0.56 nm per bp. The molecular extension of each of these
units provides a convenient reaction coordinate to study the overstretching transition. The free
energy profile of each unit along this reaction coordinate, schematically represented in Fig. S3, is
dictated by four fundamental parameters: the free energy difference between compact and extended
state $\Delta G$, the forward energy barrier $\Delta G^\ddagger_+$, the backward barrier being 
$\Delta G^\ddagger_-=\Delta G^\ddagger_+-\Delta G$, and the
distance of the transition state from both the compact state $x^\ddagger_+$ and the extended 
state $x^\ddagger_-$.
The unidirectional transition rates between the two conformations $k_+$ and $k_-$ depend on an additional
parameters $\Omega$, a kinetic pre-factor which is related to the viscous drag experienced by the molecule
in its motion along the reaction coordinate and to the shape of the molecular potential energy near
the transition state. Thus, according to Kramers-Bells theory, the $A_+$ and $A_-$ parameters are:
\begin{equation}
\label{prefactor1}
A_+=\Omega\; e^{-{\Delta G^\ddagger_+-\Delta G\over k_BT}}
\end{equation}
and
\begin{equation}
\label{prefactor2}
A_-=\Omega\; e^{-{\Delta G^\ddagger_+-\Delta G\over k_BT}}
\end{equation}
The force $F_m$ corresponding to the minimal relaxation rate $r(F)$ is
$$
F_m={-{\Delta G- k_BT \ln(x^\ddagger_+/x^\ddagger_-)\over x^\ddagger_++x^\ddagger_-}}=
$$
\begin{equation}
\label{fm}
=F_e-{{\Delta G- k_BT \ln(x^\ddagger_+/x^\ddagger_-)\over x^\ddagger_++x^\ddagger_-}}
\end{equation}
where $F_e$ is the coexistence force, the force at which the probabilities to reside in the compact and
extended state ($p_\text{cmp}$ and $p_\text{ext}$) are equal.

Also the equilibrium probabilities at each force are known:
\begin{equation}
\label{pcmp}
p_\text{cmp}(F)={1\over 1+
\exp\left(-\Delta G -F(x^\ddagger_+ + x^\ddagger_-)\over k_BT \right)}
\end{equation}
and
\begin{equation}
\label{pext}
p_\text{ext}(F)={\exp\left(-\Delta G -F(x^\ddagger_+ + x^\ddagger_-)\over k_BT \right)\over 1+
\exp\left(-\Delta G -F(x^\ddagger_+ + x^\ddagger_-)\over k_BT \right)}
\end{equation}
$p_\text{ext}(F)$ grows monotonically with $F$, having maximal first derivative at the coexistence force,
$F_e=\Delta G/(x^\ddagger_+ + x^\ddagger_-)$. At this force $\Delta G= F(x^\ddagger_+ + x^\ddagger_-)$, that is, the work done for the elongation
equals the free energy difference between the compact and the extended state. We note that only for
symmetric landscapes,
$x^\ddagger_+ = x^\ddagger_-$, $F_e$ coincides with $F_m$, the force of minimal relaxation rate.
During a positive staircase of force steps, when the force is changed from $F$ to $F + \Delta F$ the
probability to be in the extended state grows and, after equilibration, 
$N \left(p_\text{ext}(F+\Delta F)-p_\text{ext}(F)\right)$ units
have extended, where $N$ is the total number of extensible units. It is therefore straightforward to
compute the equilibrium force-extension profile which, according to the expressions for pext and
pcmp, only depends on the two parameters $\Delta G$ and $x^\ddagger_+ + x^\ddagger_-$. 
In Fig. S4 the experimental force extension
relation (black trace) from Fig. 2b is compared to the theoretical relation (red dashed line)
obtained with the  and  chosen by fitting the elongation rates with the procedure described
in the test. The agreement shows that the model developed to fit the observed relaxation kinetics is
also capable of reproducing the equilibrium kinetics.
\begin{figure}[]
\includegraphics[clip,width=6 cm,angle=-90]{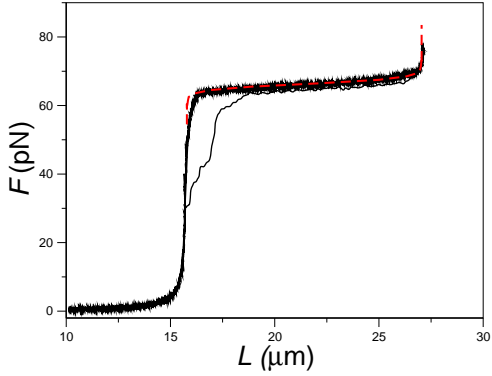}
\caption{Comparison between the experimental force-extension relation (black traces) and the
theoretical relation (red trace) calculated with the $p_\text{ext}$ and $p_\text{cmp}$ equilibrium probabilities and the
parameters $\Delta G$ and $x^\ddagger_++x^\ddagger_-$ obtained by fitting the elongation rates elicited by 
force steps. In order to reproduce the experimental curve the base pair separations of B and S state were set 
to 0.325 nm and 0.57 nm respectively.}
\label{S4}
\end{figure}

The molecular elongation at force $F$ is
\begin{equation}
\Delta L_e(F)\simeq N \partial_F p_\text{ext}(F) \Delta F (x^\ddagger_+ + x^\ddagger_-)
\end{equation}
where $p_\text{ext}$ has been expanded to the first order. An expression for $r(\Delta L_e)$ can be obtained by
inverting $\Delta L_e(F)$ and substituting into $r(F)$. $\Delta L_e(F)$, however, is a non monotonic function of $F$,
growing before $F_e$ and diminishing after. $F(\Delta L_e)$ has therefore two branches: $F_\text{low}(\Delta L_e)$ for $F< F_e$ and
$F_\text{high}(\Delta L_e)$ for $F> F_e$. Substituting the two branches into $r(F)$ one gets two expressions for $r_\text{low}(\Delta L_e)$
and $r_\text{high}(\Delta L_e)$, which, after some algebra, can be shown to diverge in the low extension regime with
exponents $x^\ddagger_-/(x^\ddagger_+ + x^\ddagger_-)$ and $x^\ddagger_+/(x^\ddagger_+ + x^\ddagger_-)$ respectively. 
More precisely
$$
r_\text{low}\simeq \Omega \left({4 L_\text{TOT} (x^\ddagger_+ + x^\ddagger_-)\Delta F\over k_B T \Delta L_e}\right)^{x^\ddagger_-/(x^\ddagger_+ + x^\ddagger_-)} 
\text{and}
$$
\begin{equation}
r_\text{high}\simeq \Omega \left({4 L_\text{TOT} (x^\ddagger_+ + x^\ddagger_-)\Delta F\over k_B T \Delta L_e}\right)^{x^\ddagger_+/(x^\ddagger_+ + x^\ddagger_-)} 
\end{equation}
where $\Delta L_\text{TOT}$ is the total elongation due to overstretching.
In our case using the values for $x^\ddagger_+$ and $x^\ddagger_-$ obtained by fitting the rate versus force relations , these
two exponents amount to 0.4 and 0.6. In Fig. 3c of the text, the logarithmic plot of the rate versus
elongation indicates that the exponent of the power equation is -0.6, the same as that expected for
$r_\text{high}-\Delta L_e$ branch, to which the majority of the data refer. This suggests that, although some degree of
interface propagation might be present in the elongation process, the cooperative elongation of
regions of DNA approximately 25 bp long represents the rate limiting step for the overstretching
transition. The downward shift of the rate-versus-elongation profile following a decrease in $\Delta F$ as
depicted in Fig. 3c also quantitatively agrees with Eq. S7 which predicts that, for the same
elongation, rates should be proportional to $\Delta F^{x^\ddagger_\pm/(x^\ddagger_+ + x^\ddagger_-)}$. 
Note that Eq. S7 predicts also that a
downward shift similar to that in Fig. 3c would be observed using 2 pN steps in a molecule $1/4$ the
length of the whole DNA molecule. In fact, for the same elongation, rates should
be proportional to $\Delta L_\text{TOT}^{x^\ddagger_\pm/(x^\ddagger_+ + x^\ddagger_-)}$, where $\Delta L_\text{TOT}$, the total overstretching elongation, is a fixed fraction (0.7) of the
length of the molecule.

\subsection*{3. Fitting the relation between rate of elongation and force with the four parameter equation}
The equation expressing the relation between the rate of elongation ($r$) and force ($F$) is:
\begin{equation}
r=k_++k_-=A_+\exp\left(F x^\ddagger_+ \over k_BT \right)+A_+\exp\left(F x^\ddagger_+ \over k_BT \right)
\end{equation}
and is fitted to the $r$-$F$ data on the entire force range for both force step sizes (purple line in Fig.
3d). The equation is a downward convex function of $F$ with a minimum at
\begin{equation}
F_m={k_BT \ln(A_-x^\ddagger_-/A_+x^\ddagger_+) \over x^\ddagger_+ + x^\ddagger_-}.
\end{equation}

The values of the four parameters are: $\ln A_+ = -34.12 \pm 1.65$, 
$x^\ddagger_+= 2.21 \pm 0.11$ nm, $\ln A_ = 59.19 \pm 3.55$; 
$x^\ddagger_-= 3.70 \pm 0.23$ nm (so that $F_m$ = 65.81), and are almost
identical to those determined by separately fitting the low and high force branch of the relations,
assessing the validity of the assumption in the text that in the two sides of the overstretching
transition the contribution of the minor term can be neglected.

\subsection*{4. Introducing heterogeneity in the model.}
In the text we have assumed that DNA overstretches in discrete, cooperative units which are all
characterized by the same size and the same free energy profile. Such assumptions are clearly an
oversimplification since it is more sensible to expect that, whatever the microscopic mechanism
generating the observed cooperativity, the sizes of the molecular regions undergoing cooperative
overstretching will cover a continuous and somewhat extended range of sizes. We here test how the
behavior of the proposed model is affected by the introduction of heterogeneity in the size of the
cooperatively overstretching units .
Monte Carlo simulations analogous to that described in the text were performed for a set of
2200 two state units, each of 22 bp. . In this set of simulations, however, a certain amount of
heterogeneity was introduced in the structural and energetic parameters $x^\ddagger_+$, $x^\ddagger_-$, 
$\ln A_+$, $\ln A_-$ and $\Delta G^\ddagger_+$
which maintained in average the values obtained by fitting the experimental data. The system was
integrated for 5 s
\begin{figure}[]
\includegraphics[clip,width=6 cm,angle=-90]{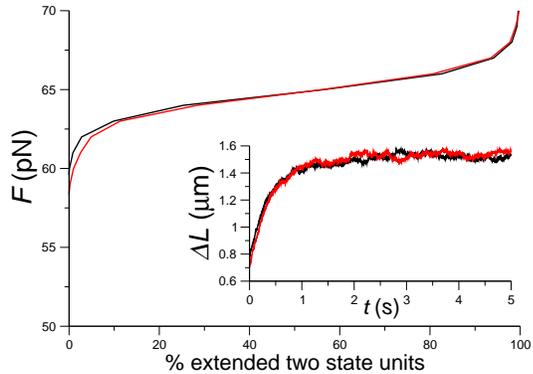}
\caption{Equilibrium force-extension relation of a system of 2200 two state units of identical size
(black) and with 30\% variation in size (red). The degree of extension is expressed as \% of extended
units. Inset: time course of elongation following 1 pN force step (intermediate between the 0.5 and
2 pN force steps of the experiments) to 65 pN; red and black as in the main frame.}
\label{S5}
\end{figure}

Fig. S5 shows that a stochastic variability of 30\% in $x^\ddagger_+$ and $x^\ddagger_-$ does not significantly alter the
profile of the transition at equilibrium. The apparent cooperativity of the transition, assessed by
means of a sigmoidal fit of the equilibrium profile, increases less than 10\% as soon as the
parameters $\ln A_+$ and $\ln A_-$ are chosen proportionally to $x^\ddagger_+$ and $x^\ddagger_-$. Also the elongation following a
force step (see inset of Fig. S5 for the elongation at 65 pN) appears to satisfactorily reproduce
the observed time course since, despite being in principle a sum of exponentials, it still fits a single
exponential (dashed line). Notably even the noise in the calculated trace is similar to the observed
one ($\pm 1$ $\mu$m).
In the previous analysis we have varied both the relevant energetic parameters - the free energy
barrier $\Delta G^\ddagger_+$ and the free energy difference between extended and compact state $\Delta G$ - 
that dictate the
relaxation rates of each cooperatively overstretching unit. By varying them independently one can
get a feeling about the possible range the variability

In order to analyze exclusively the influence in heterogeneity in the size of the cooperatively overstretching 
units on the system kinetics in the previous analysis we have varied in a coordinated fashion both the relevant 
energetic parameters - the free energy barrier $\Delta G^\ddagger_+$ and the free energy difference between extended 
and compact state $\Delta G$ . By varying them independently, instead, one can get a feeling about the effects of 
sequence variability.

\begin{figure}[]
\includegraphics[clip,width=6 cm,angle=-90]{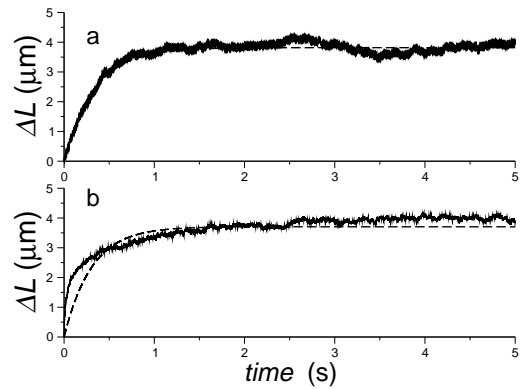}
\caption{Time course of an elongation following a 1 pN force step to 65 pN calculated for a
system composed of identical two-state units (a) and for one with a 10\% heterogeneity in 
$\Delta G^\ddagger_+$ (b).
Dashed lines are single exponential fits.}
\label{S6}
\end{figure}

The introduction of heterogeneity in the free-energy barrier $\Delta G^\ddagger_+$ is not expected to change the
equilibrium profile of the transition since the equilibrium probabilities $p_\text{cmp}$ and $p_\text{ext}$ only depend on
DG, but it does change the relaxation kinetics. Fig. S6 shows that the introduction of a Gaussian
variability in $\Delta G^\ddagger_+$
+ with standard deviation of 10\% produces an elongation time course which
departs markedly from exponential. Although an exponential fit holds very similar results for
simulations performed with a fixed and a variable free energy barrier, in this last case the time
course is characterized by the presence of an early fast rising portion corresponding to the smallest
barriers in the system. This feature is absent in the observed time courses (Fig. 3a), indicating that
any variability in the value of the free energy barrier is lower than 10

As far as $\Delta G$ is concerned, a Gaussian variability with standard deviation of 1
cooperativity by only 10\% (data not shown) without substantial effects on the elongation rate.
Larger $\Delta G$ variabilities, however, quickly introduce slow components in the elongation time course,
preventing from reaching equilibrium in the 5 s observation time. The fact that the experimental
stretching curves are almost superposable to the release curves therefore suggests that also the
variability on $\Delta G$ is lower than 10\%.

In conclusion, by interpreting our experimental data with a two state model where all cooperatively
overstretching units have the same length, we introduce only a small underestimate in the degree
of cooperativity of the system. Moreover, the exponential character of the observed kinetic profiles
and the good equilibration reached by the process in the 5 s time scale suggest that the
heterogeneity in the energetic parameters is lower than 10\%.

\end{document}